\documentclass{article}
\usepackage{emulateapj}
\usepackage{float,epsfig,psfig}
\usepackage{pstricks}

\def\hea4{{\it HEAO~A4}}
\def\heaoa2{{\it HEAO~A2}}
\def\heao1{{\it HEAO~1}}

\def\amin{$^\prime$}

\def\eg{{\it e.g.}~}

\def\h0{$H_{\rm o}=50$~km~s$^{-1}$~Mpc$^{-1}$}
\def\q0{$q_{\rm o}$}

\def\msun     {M$_{\odot}$}

\def\etal    {{et~al.}~}

\def\cms3  {~{cm$^{-3}$}}

\begin{document}

\submitted{submitted to ApJ, June 28 2001; accepted April 2 2002}
\title{{Chandra Observation of LMXBs in the elliptical galaxy M84}}  
\author{A.~Finoguenov$^{1,2}$, and C.~Jones$^2$}
\affil{
{$^1$ Max-Planck-Institut f\"ur extraterrestrische Physik,
             Giessenbachstra\ss e, 85748 Garching, Germany}\\
{$^2$ Smithsonian Astrophysical Observatory, 60 Garden st., MS 3, Cambridge,
  MA 02138, USA}}
\authoremail{alexis@head-cfa.harvard.edu}


\begin{abstract}
  
  We present characteristics of the X-ray point source population in the M84
  galaxy, observed by {\it Chandra} ACIS-S. We find an excess in the number
  of sources centered on M84, with a spatial distribution closely
  corresponding to the M84 stellar light. Given an absence of recent
  star-formation, accreting binaries are the only candidates for the M84
  X-ray sources. The majority of M84 sources (with luminosities exceeding
  $10^{38}$ ergs/s) exhibit hardness ratios expected from multi-temperature
  black-body disk emission. The most luminous sources, which we attribute to
  accreting black holes exhibit X-ray colors typical of a black body
  spectrum. We also identify the sources whose X-ray colors match the
  expectations for constituents of the Cosmic X-ray Background. The number
  of such sources agrees with that expected to be background sources. After
  correcting for incompleteness in the source detection, we find a
  $log(N)-log(S)$ for M84 similar to that of the elliptical galaxy NGC4697,
  i.e.  having a break at a luminosity of
  $L_b=2.4^{+0.6}_{-0.3}\times10^{38}$ ergs/s, approximately the Eddington
  limit on the isotropic luminosity for accretion onto a neutron star. The
  slope of the luminosity function above the break provides evidence for a
  mass distribution in the M84 accreting black holes.

\end{abstract}


\section{ Introduction}

Observation of X-ray sources in the Milky Way galaxy began with the launch
of the first X-ray detector (Giacconi \etal 1962). The study of accreting
systems at all wavelengths is now a well-developed field (for a recent
review see Tanaka \& Shibazaki 1996; theoretical progress is presented by
Iben \etal 1995; Kalogera \& Webbink 1998). With the advent of high angular
resolution X-ray observations, first with Einstein and ROSAT and now with
Chandra and XMM-Newton, it has become feasible to study the X-ray
populations in other galaxies.

\begin{figure*}
\includegraphics[width=3.6in]{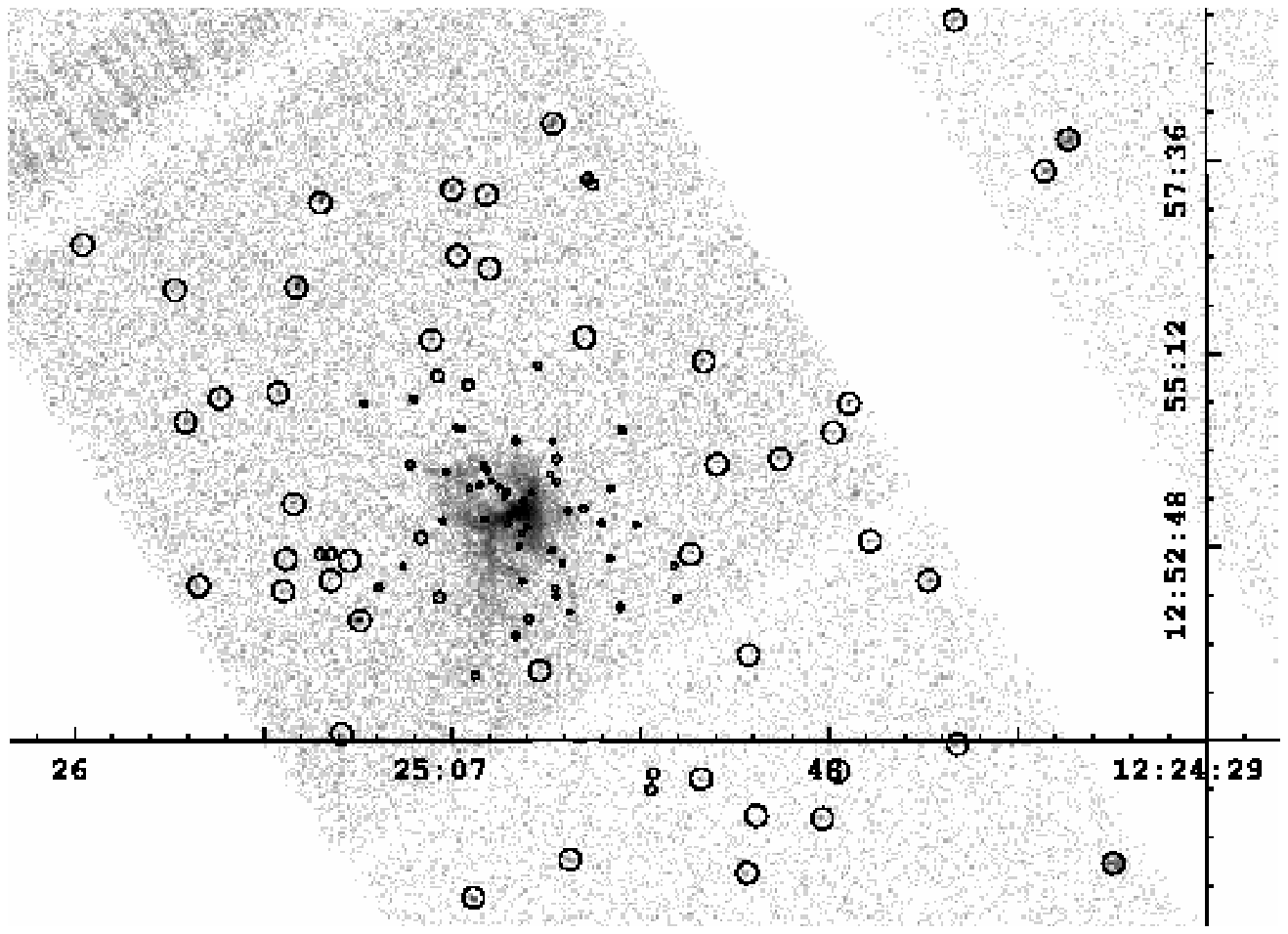}\hfill \includegraphics[width=3.3in]{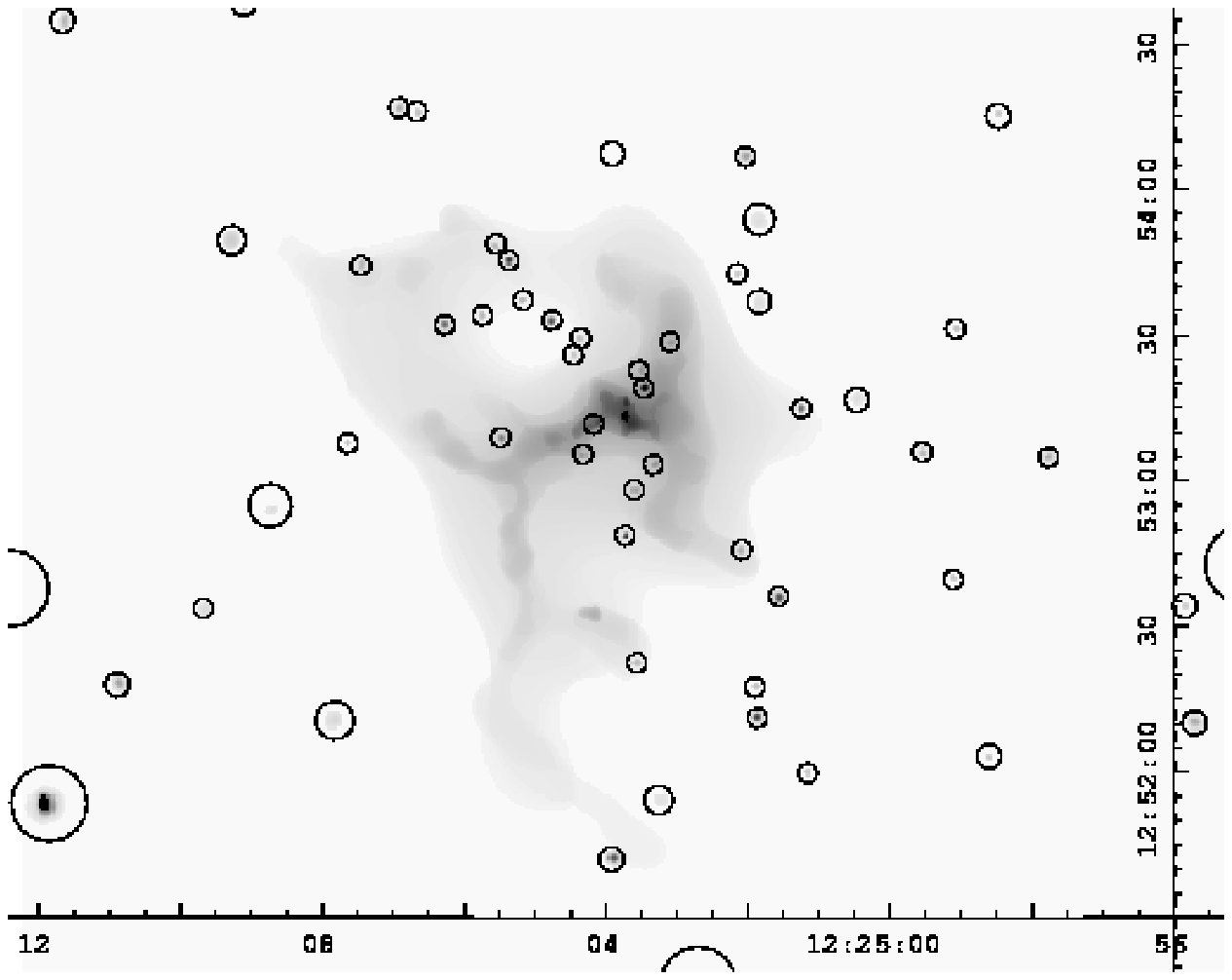}
\vspace*{0.5cm}

\includegraphics[width=3.3in]{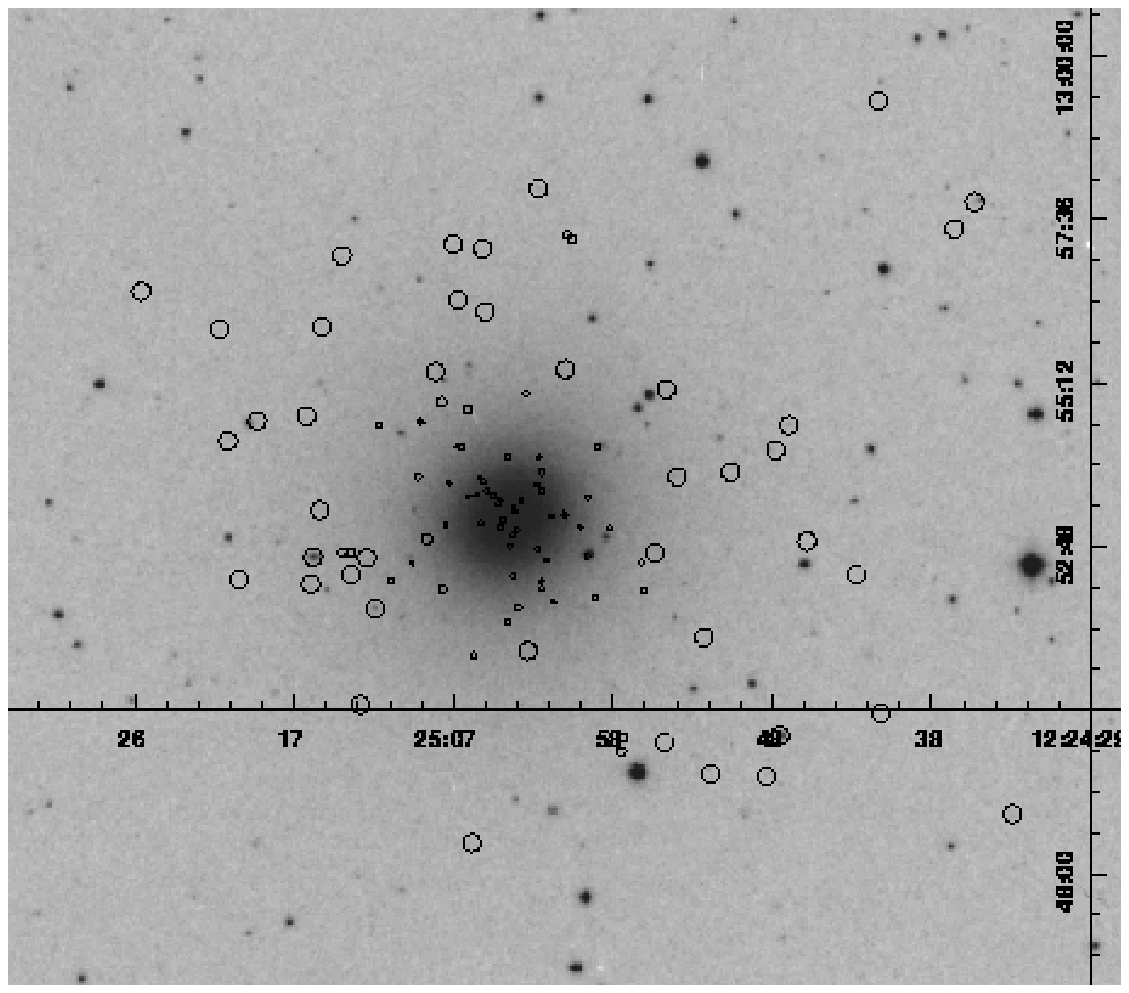}\hfill \includegraphics[width=3.7in]{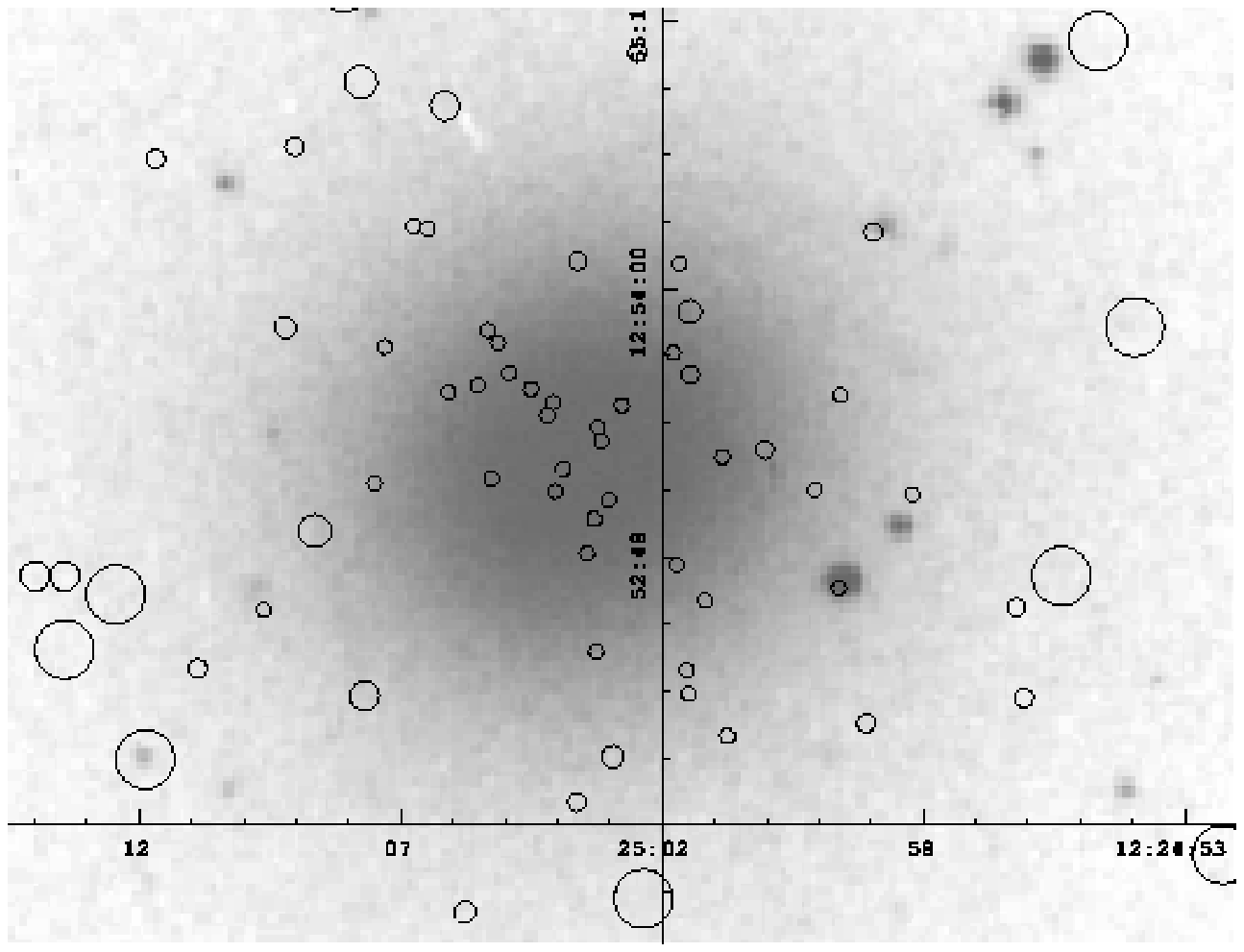}

\figcaption{ {\it Upper panels:} Chandra image of the ACIS field around M84
{\it (left)} and wavelet-decomposed image of the central part of M84 {\it
(right)}, used to calculate the source surface density. Circles mark
individual point sources. The energy band is 0.4--5.0 keV. {\it Lower row:}
DSS (optical) image of M84 overlaid with X-ray point source positions.
\label{fig:north}
}
\end{figure*}

This {\it Paper} presents a detailed study of Chandra observation of the
point sources in M84, an elliptical galaxy in the core of the Virgo cluster
(study of the diffuse X-ray emission of M84 is presented in Finoguenov \&
Jones 2001). The particular importance of population studies in early-type
galaxies comes from the small number of possible avenues leading to the
X-ray source production. As most ellipticals are old stellar systems, only
accretion of matter in low mass X-ray binaries (LMXB) can provide enough
power to explain X-ray luminosities that exceed $10^{37}$ ergs/s. The
question of how low is the mass of the secondary star is closely connected
to studies of the stellar population (\eg\ Worthey 1994) and recent merger
activity (\eg Schweizer \& Seizer 1992). The latter can be responsible for
differences in the X-ray population in the centers of galaxies, compared to
that in the halos.  The dependence of LMXB activity on star-formation rate
is considered in White \& Ghosh (1998).

When one star in a binary system becomes a compact object (primary), the
accretion process can be triggered through the Roche-lobe overflow of
material from the evolved low-mass companion star (secondary). Another
mechanism for triggering accretion at epochs long after the cessation of
star-formation, is by shrinking the binary star separation through magnetic
braking (outward transport of the angular momentum in magnetic stellar
winds, Verbunt \& Zwaan 1981) or gravitational radiation losses (Paczy\'nski
1967) leading to accretion even from a main sequence star. Accretion onto a
white dwarf is identified through a super-soft X-ray spectrum (\eg\ Kahabka
\& van der Heuvel 1997). However, our observation is not sensitive to this
mechanism, since detecting white dwarfs in our softest energy band ($0.3-1$
keV) would imply a luminosity for the source (calculated assuming a spectral
temperature of 35 eV), far exceeding the corresponding Eddington limit.

The luminosity function of X-ray sources in M84 can be expressed as a
convolution of the accreting binary mass-function and X-ray luminosity
function for equal mass primaries. Neutron stars, the most numerous class of
accreting binary primaries, have very similar masses, 1.4 \msun. Thus, the
source luminosity function below the Eddington luminosity for a neutron star
($1.8\times10^{38}$ ergs/s) is dominated by the X-ray luminosity function of
neutron-star LMXB (NS LMXB). Their luminosities are proportional to the
accretion rate determined by the orbital period and mass loss (Wu
2001). Most of the known black-holes in binary systems have a mass of 6
\msun\ (Tanaka \& Shibazaki 1996), however recent studies indicate a more
complicated black hole mass function (Bailyn \etal 1998; Fryer \& Kalogera
2001).

In this {\it Paper} we study the spatial distribution of LMXB in M84
(\S\ref{sec:dist}) and derive the integral $log(N)-log(S)$ distribution
(\S\ref{sec:logn}). We study source hardness ratios and conclude with a
discussion on the emission mechanism for LMXB
(\S\ref{sec:hr}). Interpretation of the M84 LMXB luminosity function is
given in \S\ref{s:d}.

\section{Analysis of the ACIS observation}

In the ACIS-S 28.7 ksec observation of M84, point source detection was done
in the 0.4--5.0 keV energy band, where the source signal-to-noise is
highest. Throughout sections \S\ref{sec:dist}--\S\ref{sec:hr}, luminosities
are cited for the 0.4--10 keV band, using the countrate-to-luminosity
conversion for the best-fit power law photon index of 1.4, derived for the
combined spectra of all point sources within two effective radii (Finoguenov
\& Jones 2001). Given the duration of the observation, and assuming a 17 Mpc
distance to M84, one ACIS-S count corresponds to $10^{37}$ ergs/s. Estimates
of the detection rate for the background sources were made using a power law
photon index $\Gamma=1.7$ and the $log(N)-log(S)$ determination by Giacconi
\etal (2001).

Source detection was carried out using the matched filter (wavelet)
technique (Vikhlinin \etal 1995). We performed simulations to estimate the
detection efficiency of this method on our field, as described below
(\S\ref{sec:logn}). In addition we estimate the effects of variations in the
PSF on the resulting signal-to-noise and thus source detection
efficiency. Variations in the PSF cause the apparent source size to increase
with increasing off-axis angle. Thus, we run the detection procedure twice,
first for the central $10'$ radius region, where we use a full resolution
ACIS pixel image ($0.492''$ on a side) and second employing a 4 by 4 pixel
binning for the full $15.6'\times11.1'$ field, shown in Fig.\ref{fig:north},
which includes S3 and portions of S2, I2 and I3 chips. At large radii, a
decrease in the detection efficiency arises due to the larger detection
region for the source counts and therefore for the background. We merged the
source lists from the two methods, thus removing duplications. To determine
the source count rate, we calculate exposure maps, accounting for the
spatial non-uniformity of the CCD quantum efficiency. We take vignetting
(weighted by the mean spectrum of sources, $\Gamma=1.4$) and dithering into
account, using CIAO contributed software.

\section{Spatial distribution of X-ray binaries in M84}\label{sec:dist}

Before we proceed to describe our results, we note that for M84, we can
study the X-ray population at a few effective radii, while strong diffuse
emission prevents us from detecting faint sources very close to the galaxy
center.

The Chandra ACIS image of the M84 field is shown in Fig.\ref{fig:north}. We
show X-ray images of the full field and the center of M84 with detected
sources marked with circles. The enhanced background (in the upper left of
the top left image) is from a neighboring CCD chip (S4), which we omitted
from analysis.

Of the 106 sources shown in Fig.\ref{fig:north} we exclude 3 sources
associated with foreground objects from all subsequent analysis. We
identified these as corresponding to an optical counterpart of size or
luminosity exceeding that expected for globular clusters at the M84
distance. Spatial analysis of the source density does not reveal any
peculiarities at the 5$\sigma$ level. At the 3$\sigma$ significance level
(Fig.\ref{fig:2dmap}), a clustering of sources on sub-arcminute angular
scale is seen, as well as the effect of non-uniform source detection in
regions with strong diffuse emission, corresponding to the $\cal H$-shape of
the hot gas in M84, seen in the upper panels of Fig.\ref{fig:north}.  Strong
clustering on sub-arcminute angular scales is an attribute of the sources
that constitute the Cosmic X-ray Background (\eg\ Vikhlinin \& Forman 1995).

\includegraphics[width=3.2in]{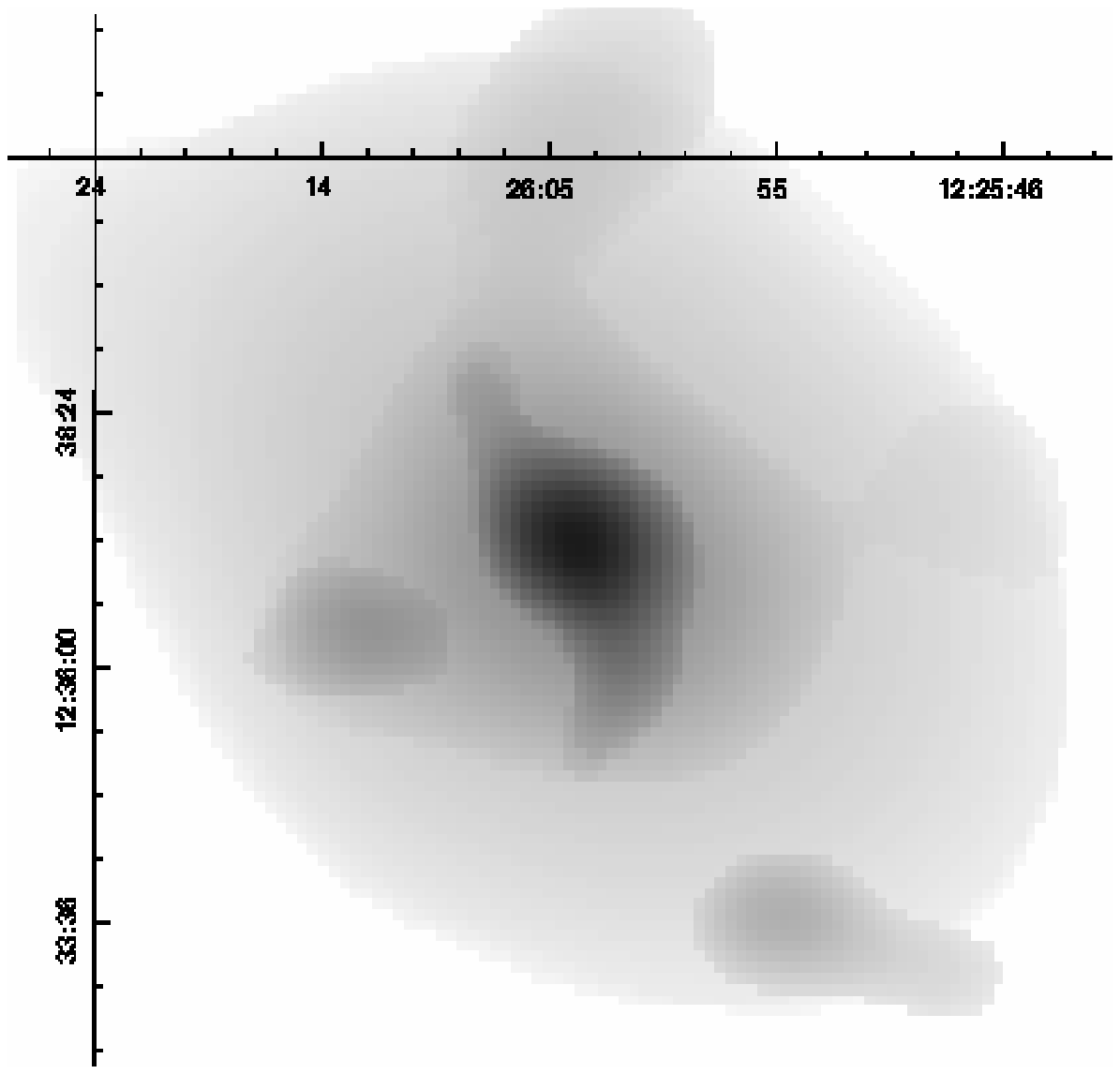} 

\figcaption{Map of the surface source number density in M84. The major
concentration of sources is due to LMXBs following the light of the galaxy
and is centered on M84. Deviations from the symmetry around the center are
caused by variation in the detection threshold due to strong diffuse
emission as well as contributions from the CXB sources, which exhibit
pronounced source clustering on sub-arcminute angular scales.
\label{fig:2dmap}
}

\includegraphics[width=3.2in]{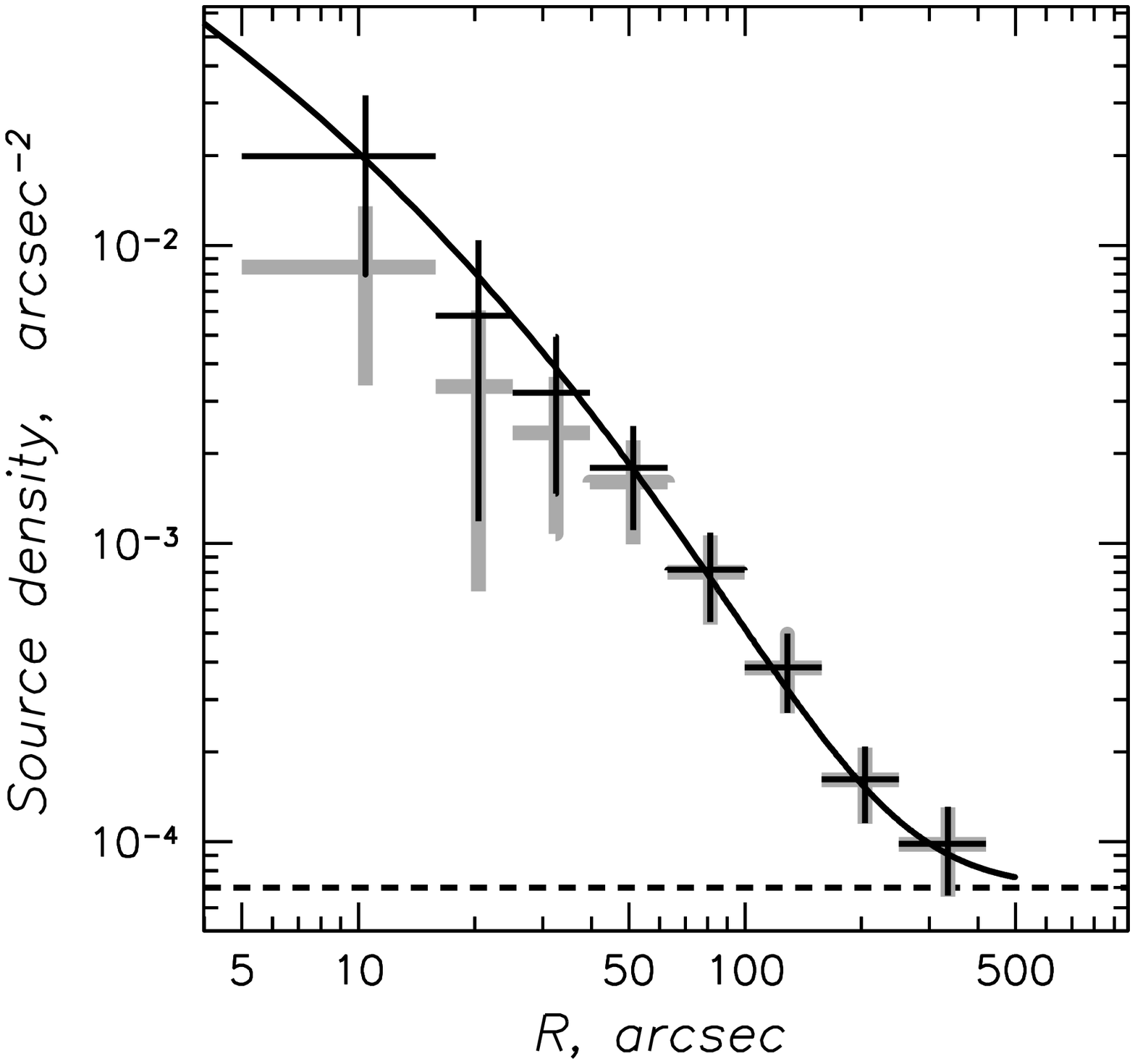} 

\figcaption{Source number density profile. The solid line denotes the galaxy
light profile. The black (grey) crosses denote the distribution of Chandra
sources corrected (uncorrected) for reduced detection efficiency due to the
presence of the diffuse source. A dashed line shows the expected level of
background X-ray sources (Giacconi \etal 2001).
\label{fig:pro}
}

The profile of the source number density, centered on M84, is presented in
Fig.\ref{fig:pro}. While the outer part is described well by the stellar
light distribution, taken as a de Vaucouleurs (1/4) law with
$r_e=1^{\prime}$ (where $r_e$ is the effective radius, within which the
galaxy contains half of its light), the center reveals a deficit of sources,
due to the high detection threshold, caused by the bright diffuse X-ray
emission. In the next section, we develop a method to correct for this
effect, with the results also shown in Fig.\ref{fig:pro}.

With the normalization shown in Fig.\ref{fig:pro}, and assuming a distance
to M84 of 17 Mpc (the corresponding luminosity in the B band is
$L_B=4.5\times10^{10} L_{\odot}$ within $r_e$), the LMXB rate per $L_B /
10^{10} L_{\odot}$ is $23\pm2$ for X-ray luminosities $>3\times10^{37}$
ergs/s. The fact that the X-ray binary distribution is similar to that of
the optical light, strongly suggests a similar formation history for single
and binary stars. Latter mixing should not play a strong role, since
metallicity gradients are detected in this galaxy (Kobayashi \& Arimoto
1999), and otherwise would be erased.

\section{$log(N)-log(S)$}\label{sec:logn}

\includegraphics[width=3.2in]{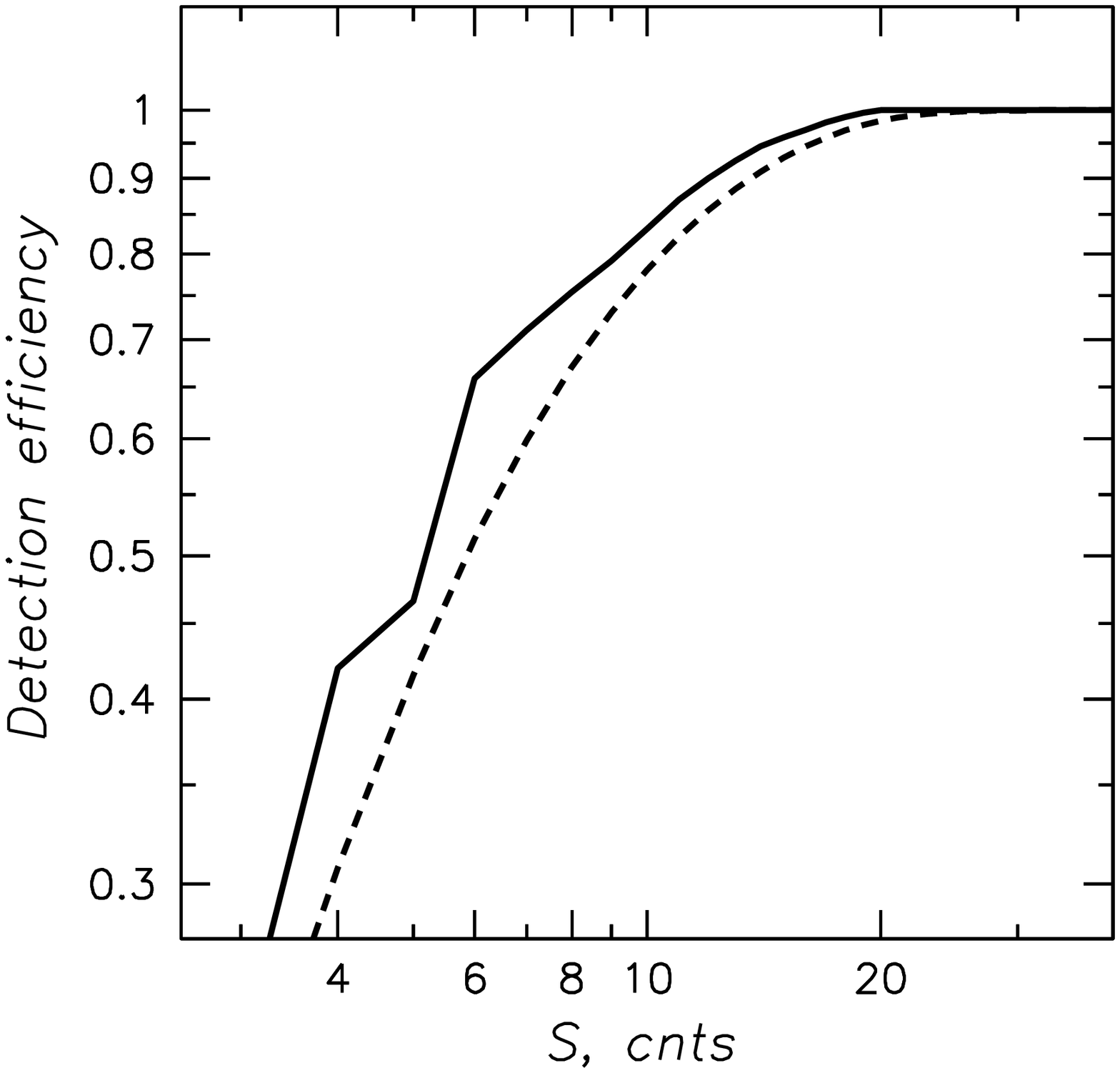}

\figcaption{Efficiency of source detection as a function of detected
counts. The solid line illustrates the decrease in the detection efficiency,
seen at source counts less than 20, caused by the bright diffuse X-ray
emission, changes in sensitivity (mostly due to quantum efficiency
differences between the chips) and the reduced signal-to-noise due to PSF
degradation at large off-axis angles. The dashed line indicates the
resulting efficiency due to the Poisson migration.
\label{fig:eff}
}

In deriving the efficiency for source detection, we simulated the effect of
the diffuse X-ray emission on the resulting signal-to-noise requirements for
our source detection at a $5\sigma$ statistical threshold. We also took into
account the spatial distribution of sources, which are concentrated toward
the center, where the 'background' is higher. This further reduces the
detection sensitivity. These effects are important for source detection
close to the center of M84. For sources outside the effective radius of M84,
changes in the sensitivity (due primarily to quantum efficiency differences
between CCD chips) and the reduced signal-to-noise, due to PSF degradation
at large off-axis angles, play important roles in the detection. We estimate
the resulting detection efficiency in Fig.\ref{fig:eff}. The effects of
strong diffuse emission are important for sources with fewer than 15 counts,
while changes in the PSF affect sources with 6 counts or fewer. Because of
the Poisson noise, the shape of the source luminosity distribution will be
further modified (Poisson migration) and will be different for different
exposure times. While a proper reconstruction requires a response matrix
with the luminosity resolution defined by a Poisson process, in the present
analysis, we simply correct for the amplitude of the effect, by convolving
the detection efficiency with the Poisson probability distribution and show
the resulting detection efficiency (which is equivalent to the completeness)
in Fig.\ref{fig:eff}. In the following derivation of the $log(N)-log(S)$, we
will correct for this incompleteness in the survey. Finally, changes in the
sensitivity (effective area and quantum efficiency) are most important for
flux correction and affect the resulting slope at high fluxes.

The resulting $log(N)-log(S)$ is shown in Fig.\ref{fig:logn}. In further
analysis we accounted for the contribution from extragalactic sources
(Giacconi \etal 2001). There is a break at the level of 19 sources (with an
expected CXB contribution of 6 sources) at a corresponding luminosity of
$L_b=4.6^{+1.2}_{-0.6}\times10^{38}$ ergs/s (errors quoted are 68\%
confidence limits). The integral slope at lower fluxes is $-0.87\pm0.07$,
while at higher fluxes, it is $-1.8\pm0.5$. The differential slopes at lower
and higher fluxes are $-1.79\pm0.18$ and $-2.7_{-1.1}^{+0.6}$,
correspondingly.  Our results for the luminosity of the break and the slope
at high fluxes are in remarkable agreement with findings by Sarazin \etal
(2000) for another elliptical galaxy, NGC4697,
($L_b=3.2^{+2.0}_{-0.8}\times10^{38}$ ergs/s and differential slopes at
lower and high fluxes of $-1.29_{-0.36}^{+0.49}$ and $-2.76_{-2.0}^{+0.8}$,
correspondingly). However, as we will argue below, luminosities of the break
in both galaxies are derived for the averaged spectrum, which is {\it not
typical} for NS LMXB emitting close to the Eddington.

Flattening of the source luminosity function below the luminosity of
$5\times10^{37}$ ergs/s, seen in NGC4697 and M31 also may be present in M84,
but since we have only one point there, derived from our detection of
sources with only 3 and 4 counts, it would certainly be an
over-interpretation to discuss this in more detail.

\includegraphics[width=3.2in]{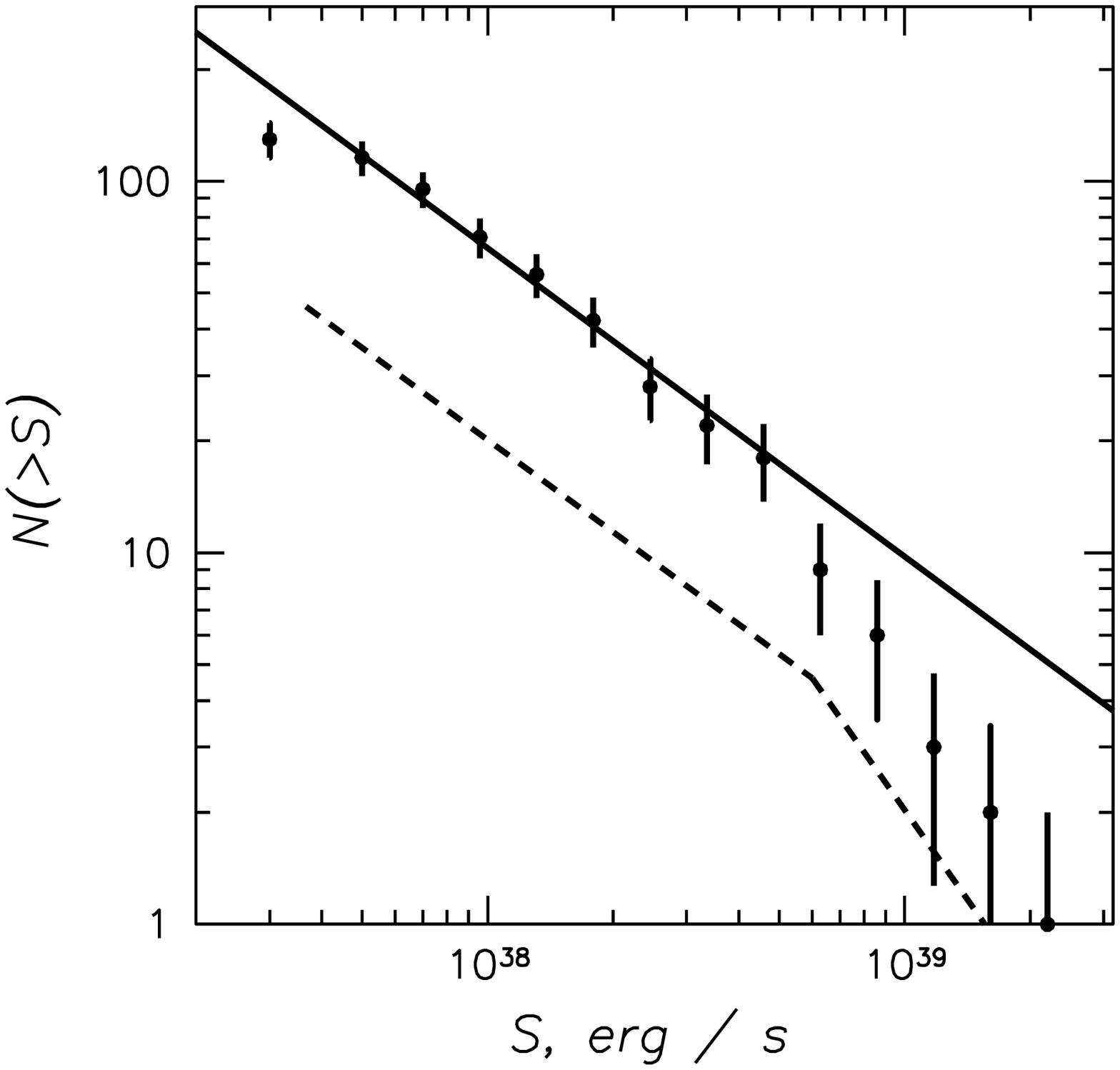}

\figcaption{$log(N)-log(S)$ of M84 sources. The dashed line indicates the
CXB. Foreground sources were removed from the source list. Solid line
indicates the fit to low fluxes. 
\label{fig:logn}
}

\begin{figure*}
\includegraphics[width=3.4in]{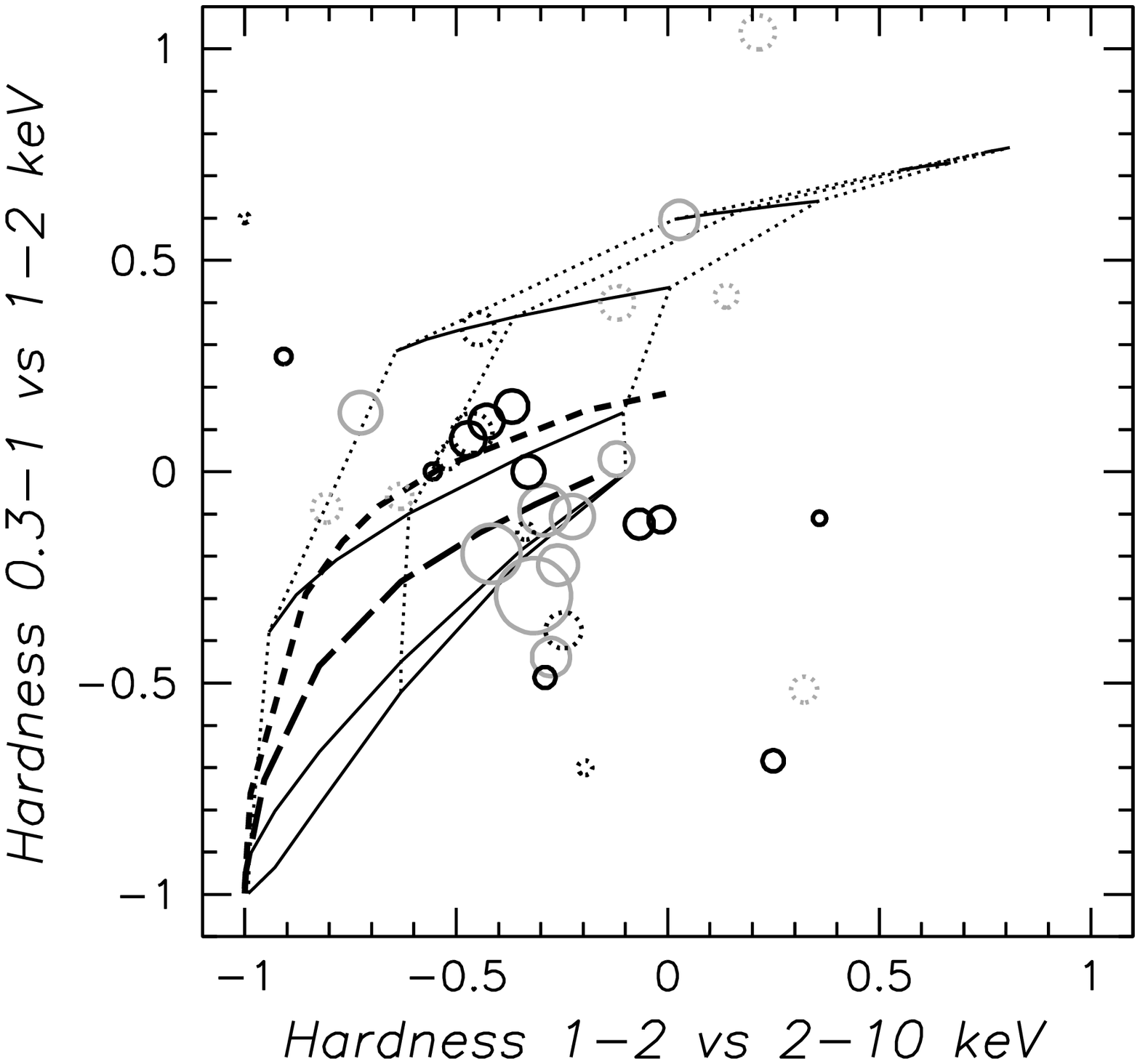}\hfill \includegraphics[width=3.4in]{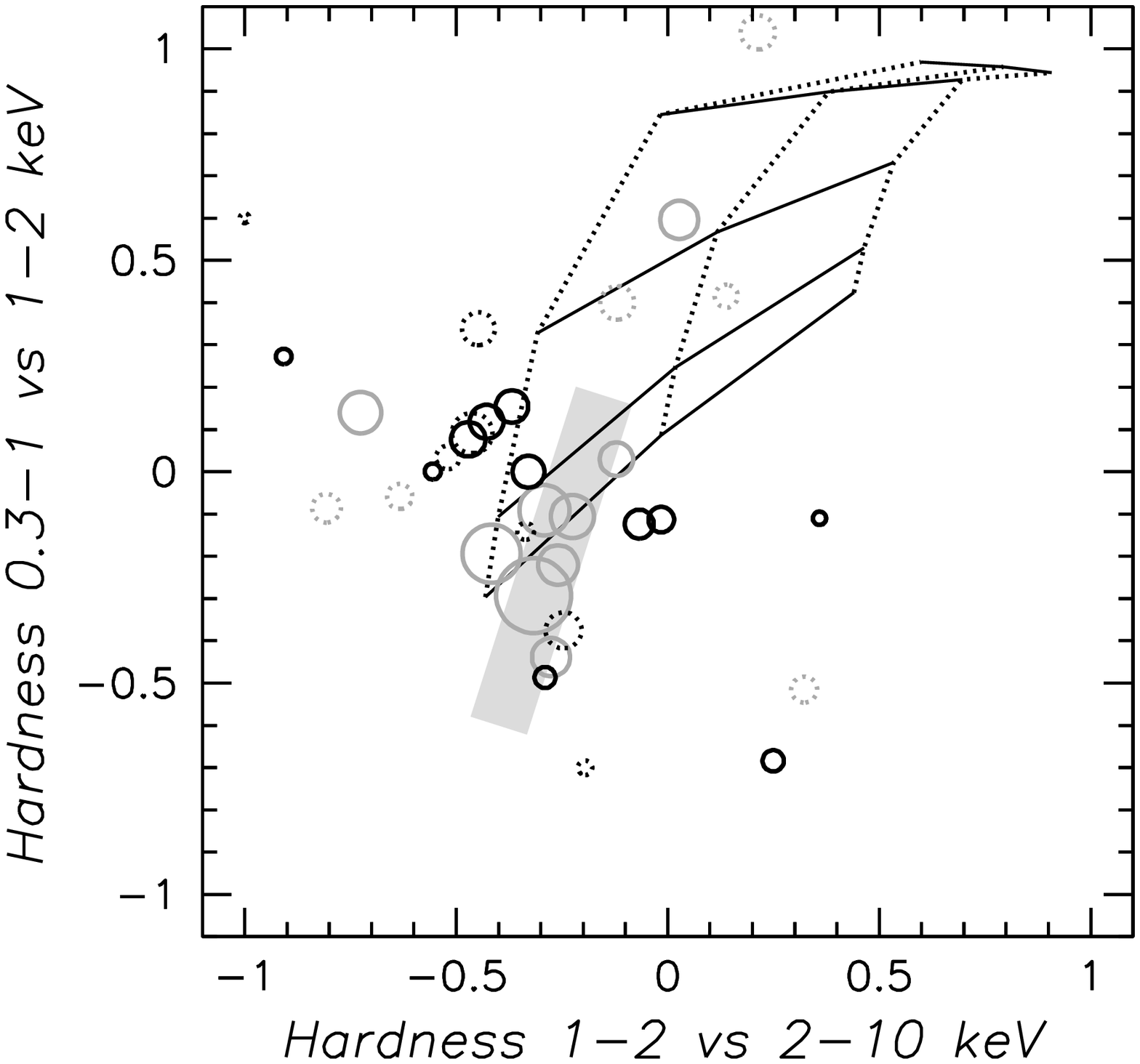}

\figcaption{X-ray spectral diagnostic diagrams, based on source hardness
ratios, determined using the background subtracted source counts in the Hard
and Soft bands as $(H-S)/(H+S)$. Data points are shown as black circles for
sources within 2\amin\ of the M84 center and as grey circles otherwise. The
size of the circle is proportional to the logarithm of the flux and circles
are drawn with a solid line when the uncertainty in the hardness ratio
estimation is less than 0.1, with a dotted line for uncertainties less than
0.2. The differences between the {\it left} and {\it right} panels are only
in the model grids. On the {\it left} the long-dashed line denotes a
hardness ratio from the bremsstrahlung model, the short-dashed line
corresponds to the model of a multicolor black-body disk by Makishima \etal
(1986). We also plot a grid corresponding to a comptonized on 10 keV plasma
black-body spectrum with a blackbody temperature of 0.01, 0.1, 0.2, 0.4,
0.8, 1.6, 3 keV (the solid lines) and optical depth for comptonization of
0.1, 1.6, 6.4 (the dotted lines). In the {\it right} panel the shaded area
denotes the loci for type-1 AGN. A grid illustrates the hardness ratios for
power law models with photon indices $\Gamma=0$, 1, 2 (dotted lines) and
neutral hydrogen absorption (in the observed frame) of $log(nH)=20.5$, 21,
21.5, 22, 22.5 (solid lines), typical for the hard background sources found
by XMM (Hasinger \etal 2001).
\label{fig:hr}
}
\end{figure*}

\section{Hardness ratios}\label{sec:hr}

In the absence of a variation in absorption across M84, plus the lack of
neutral gas in ellipticals, the source hardness ratios are particularly
straightforward to interpret.

In Fig.\ref{fig:hr} we present the hardness ratios for the 32 sources with
the best statistics, determined using the source counts in the Hard and Soft
bands as $(H-S)/(H+S)$, using three independent energy bands, $0.3-1$, $1-2$
and $2-10$ keV. Given the precision of the hardness ratio estimates, a
combination of first vs second (HR21), and second vs third (HR32) reveals
the largest spread in the points and therefore provides better tests for the
predictions from various spectral models.

We subdivide the source sample into two parts, within or outside two
effective radii, as we expect almost no background sources to be detected in
a small region close to the center of the galaxy. The larger number of
sources available at larger radii, allows us to better sample the X-ray
colors. The $log(N)-log(S)$ of M84 sources and of cosmic X-ray background
sources are quite similar in the observed flux range.  In the whole field
used for source detection, CXB sources constitute almost 1/3 of the detected
sources, i.e. in Fig.\ref{fig:hr}, 11 background sources should appear,
including 6 of high statistical significance (solid circles). Given the
energy band of Chandra and using the results of point source hardness ratios
and $log(N)-log(S)$ analysis from the deep XMM pointing of the Lockman hole
(Hasinger \etal 2001), we conclude that 1/3 of the background sources should
exhibit hard colors (within the power law -- nH grid shown in
Fig.\ref{fig:hr}, note that a variation of nH is intrinsic to AGN), while
the other 2/3's should be within the grey-shaded area on the right panel of
Fig.\ref{fig:hr} of ROSAT-type background sources (type 1 AGN). In fact a
fraction of the ROSAT-type background sources should be larger, due to
differences in the sensitivity between Chandra and XMM at high energies.

We find two concentrations of hardness ratios. The first (black circles in
Fig.\ref{fig:hr}) corresponds to that expected for a multicolor disk
black-body model (Makishima \etal 1986; {\it diskbb} model in XSPEC) with a
central black-body temperature near 0.5 keV. The second (gray circles in the
shaded area on the right panel of Fig.\ref{fig:hr}) corresponds to power law
indices between 1 and 2. The empirical relation between the luminosity and
the spectral shape, derived for the Milky Way sources, implies that luminous
high state LMXB's should be dominated by a thermal component, while for
sources in the low state, the spectrum is harder. This is in a good
agreement with our detection of a number of sources with thermal disk
emission, which includes most of the sources with flux levels near the break
in the luminosity function. Thus, although the disk black-body model cannot
account for all the observed X-ray colors, the large concentration of M84
sources about the prediction of this model allows us to distinguish them
from CXB sources. Among the remaining sources there are four that follow the
prediction of the black-body spectrum (the furthest to the left dotted line
in the comptonization grid in the left panel of Fig.\ref{fig:hr}). Their
luminosities exceed the Eddington limit for a neutron star. Other sources
exhibit both a soft-excess and a hard tail (these sources fall in the lower
right of the diagram). The hard tail in M84 sources could be produced by
comptonization on the hot plasma ($kT\sim10$ keV) with typical $\tau=1-3$
and a soft input spectrum $kT<0.1$ keV (Sunyaev \& Titarchuk 1985; Titarchuk
1994; {\it comptt} model in XSPEC).

Our conclusions regarding the origin of X-ray colors in M84 LMXBs are in
agreement with a much more developed scheme for X-ray binaries in the Galaxy
(e.g. Done 2002). The next step would be then to employ the
luminosity-spectrum relation, established for the Milky Way sources to
separate the NSs from BHs in M84 at low luminosities. Given that the
luminosities of M84 sources with well measured colors exceed $10^{38}$
ergs/s, the soft component should still be dominant in LMXBs with a
black-hole primary, yet their Milky Way prototypes exhibit a large spread in
colors. On the other hand, the majority of LMXBs with NS-primary should be
Z-sources (like Cyg X-2), that are characterized by a modest variation in
colors. Following these guidelines, X-ray black holes comprise up to 1/3-1/2
of the X-ray sources in M84. Chandra observations of the Sombrero galaxy
(Delain et al. 2002), indicates a somewhat smaller, yet similar fraction of
accreting black holes (1/4). While small-number statistics prevents us from
a firm conclusion, the observed differences are in line with the suggestion
of a gradual built-up of black-holes in the 1.5-3 \msun\ range (Prokhorov \&
Postnov 2001).

\section{Discussion}\label{s:d}

\subsection{The break: Eddington limit on isotropic radiation from
  accretion of matter on the neutron star.}


We note in \S\ref{sec:hr}, that sources near the break of the luminosity
function have similar spectra. Their spectra are softer than the average
one. Therefore, using a softer spectrum we find the break in the luminosity
function occurs at $L_b=2.4^{+0.6}_{-0.3}\times10^{38}$ ergs/s (a factor
0.52 lower than the estimate using the averaged spectrum). The Eddington
limit also should be corrected for the $\mu_p=<A>/<Z>$, which for the
typical solar metallicity of stars in M84 is 1.17. Thus the corrected value
for the isotropic emission from accretion of matter of solar elemental
composition onto a neutron star is $2.1\times10^{38}$ ergs/s, which agrees
well with our measurements of the break.

The existence of a limiting luminosity arising from accretion onto a neutron
star has been debated by several authors. Paczy\'nski \& Wiita (1980)
proposed that at supercritical accretion rates, the resulting luminosity
could be much higher than the Eddington luminosity. Thus, our result of a
very close match between the break luminosity and the Eddington luminosity
is important in modeling the structure of the inner accretion zone. Inogamov
\& Sunyaev (1999) find that at accretion rates close to critical, emission
from the surface of the neutron star becomes softer, comparable to the
emission from the disk. Compared to emission from the disk, spectrally
harder emission from the surface of the neutron star is expected at lower
accretion rates and together with black holes is probably responsible for
the harder colors of the combined spectrum.

A prediction of a cut-off in the luminosity function is not unique to the
Eddington limit. For example, on a basis of LMXB evolution, Wu (2001)
predicts a gradual decrease in the number of accreting systems with high
mass transfer rates with galactic age, with a luminosity function exhibiting
an exponential cut-off. However, a close correspondence of the break in the
luminosity function to the Eddington luminosity, also seen in star-forming
galaxies (K\"ording, Falcke, Markoff 2002), is a strong argument in favor of
the relation of the break to the limiting luminosity.

\subsection{The high end of the luminosity function: the mass function of
  accreting black holes}

Chandra observation of M84 reveal 5 to 10 sources (the number is sensitive
to the behavior of the luminosity function just above the break), whose
luminosity exceeds the Eddington limit for accretion onto a neutron star.
When BH LMXBs (LMXBs with a black hole primary) are used to explain the
high-luminosity end of the source population in M84, a problem arises in
explaining why such a steep dependence of source number on luminosity
becomes flatter below $\sim 2\times10^{38}$ ergs/s. One can seek an
explanation in terms of the specific BH LMXB luminosity function\footnote{We
define the specific luminosity function, as a distribution of luminosities
(resulting from a distribution of the accretion rates) for a {\it given}
mass black hole.}, but such a coincidence with the Eddington limit is
unlikely. On the other hand, it is natural to attribute this to the
black-hole mass-function. In this case, the differential slope of the black
hole luminosity function should be shallower than -1.79, or no break would
be seen, while a steeper slope of the luminosity function above the break is
a result of convolution of the black hole mass and luminosity
functions. Using our measurements, we can constrain the differential slope
$\alpha$ of the mass function for accreting black holes in M84 to be
$-2.7<\alpha<-0.9$, where the lower limit comes from an assumption that
black holes shine at their Eddington limit and the upper limit corresponds
to similar luminosity functions of LMXBs with black-hole and neutron star
primaries. The statistical significance of the slopes quoted above is the
same as for the slope at high luminosities (${+0.8 \atop -2.0}$).

Besides accreting black holes, explanations for the appearance of
super-Eddington sources, based on an accreting neutron star could be:
beaming of the radiation, flaring and super-Eddington accretion rates.  Each
of these scenarios has difficulties explaining the observed luminosity
function of M84. Observational examples of super-Eddington accretion rates,
such as for SS433, demonstrate that the luminosity does not exceed the
Eddington limit by more than a factor of two. Occurrence of a break in the
observed luminosity function at the Eddington limit for {\it isotropic}
radiation would be hard to reproduce in the beaming scenarios. As the
existence of stellar mass black-holes is not disputed, our conclusion on the
effect of black-hole mass function provides most likely explanation for the
observed behavior of the M84 source luminosity function at high fluxes.

The brightest candidate galactic source in M84 has a luminosity of
$1.6\times10^{39}$ ergs/s. In the inner 2\amin\ region of M84, where most of
the galactic sources should be located and where the CXB contamination is
low, the brightest source has a luminosity of $0.9\times10^{39}$
ergs/s. This is contrary to the very luminous sources found in the Antennae
galaxies and other star-forming galaxies, whose spectral characteristics
also are different (Fabbiano, Zezas, Murray 2001; Makishima et
al. 2000). Interpretation of the Antennae sources could be either via
beaming (King \etal 2001; K\"ording, Falcke, Markoff 2002; Zezas \& Fabbiano
2002) or exceeding the Eddington limits (Begelman 2002).



\subsection{Low end of the luminosity function: sampling the initial
  distribution of binaries with NS primaries}

To relate the observed luminosity function to the distribution of objects
with different physical parameters, we need to make an assumption about the
origin of the accretion process. Two major scenarios are possible, accretion
from a main sequence star or accretion caused by nuclear evolution of the
secondary.

The first scenario dominates for LMXBs in the Milky Way and can accommodate
both the high accretion rates and short periods of the LMXB systems. To
apply this model to the observed luminosity function of LMXBs in M84, we
need to consider systems with relatively high accretion rates (typically in
excess of $10^{-9}$ \msun\ yr$^{-1}$). At these accretion rates, assuming
they are persistent, observed LMXBs must have formed recently (less than a
Gyr ago). The old age of the stellar population (12 Gyr) implies that the
fraction of active systems is 1/100--1/10 of the integral LMXB production (a
factor of $\Delta T / T$). Yet, as the time since star formation increases,
different mechanisms come into play. However, for accretion from a main
sequence star, magnetic braking (Verbunt \& Zwaan 1981) can change the
behavior of period over time. For example, for a similar mass range, which
we take as $0.3-1$ \msun\ for the secondary, and using the prescription for
the magnetic braking from Wu (2001), at 12 Gyr after star-formation,
binaries with longer initial periods (0.3-1.3 days) are important, compared
to those with periods shorter than 0.3 days that dominate the accretion
process only 0.1 Gyr after star-formation. Therefore, the LMXB population in
galaxies of different ages essentially samples different parts of the
initial period distribution. The effect should be seen in comparison studies
between galaxies and will provide clues to binary formation (Kalogera \&
Webbink 1998).


For LMXBs with an evolved secondary, the luminosity and period of the system
are not independent. Assuming a 20\% efficiency for accretion, the
luminosity of LMXBs is given by $0.4\times10^{37} P_{d}^{0.93}M_2^{1.47}$
ergs/s for accretion from an evolved companion, where $P_{d}$ and $M_2$ are
the period of the binary in days and mass of the secondary in \msun\
(Webbink \etal 1983). Given the typical solar mass for the secondary, the
luminosity of $10^{38}$ ergs/s requires periods on the order of ten
days. Most LMXBs in the Milky Way have short periods, but the Milky Way
star-formation also could be recent (note that the observational appearance
of LMXBs in the Milky Way bulge is generally attributed to high kick
velocities, \eg\ White \& Ghosh 1998). While studies of periodicities of
LMXBs in ellipticals are the most direct way to disentangle systems with
evolved secondaries from those with main sequence secondaries, the required
collecting area for such observations will only be available with new
generations of X-ray telescopes (\eg\ XEUS). Element composition of the
accreting material from an evolved secondary is dominated by helium. This
increases the Eddington limit by a factor of two. Therefore, observing a
lower value for the Eddington limit, favors accretion from main sequence
stars in LMXBs.

Finally, we would like to comment on the production of LMXBs in globular
clusters. Stellar trapping by a compact object in globular clusters can lead
to LMXB formation (Kuranov, Postnov, Prokhorov 2001). Although the number of
X-ray sources associated with globulars exceeds the average expected for
their light, the number is still a small fraction of the total number of
sources ($\sim10$\%) in some ellipticals (Sarazin \etal 2000; Kraft \etal
2001). Therefore, if globulars are a primary site for LMXB production in
early-type galaxies, to explain the LMXBs outside globular clusters either
globular clusters have been disrupted or a ``kick'' expelled the LMXB (as
during the collapse of the white dwarf into the neutron star). Indirectly
the importance of globulars can be estimated by comparative studies (White
2001). In addition, the outskirts of brightest cluster galaxies, for example
NGC1399 and M87, are characterized by an astonishingly high frequency of
globulars, which explains a strong association of globulars with X-ray
sources in the {\it outskirts} of NGC1399 found by Angelini, Loewenstein,
Mushotzky (2001).


\section*{Acknowledgments}

This work was supported by NASA grants GO0-1045X and AG5-3064 and the
Smithsonian Institution. AF has benefited from discussions with Andrei
Beloborodov, Eugene Churazov, Guenther Hasinger, Vicky Kalogera, Wolfgang
Pietsch, Konstantin Postnov, Juri Poutanen, Sergey Sazonov, Rodrigo Supper,
Yasuo Tanaka and Andreas Zezas at various stages of the presented
research. AF acknowledges receiving the Max-Plank-Gesellschaft Fellowship.
Authors thank the referee for useful comments on the manuscript.

\end{document}